\documentclass[twocolumn,showpacs,superscriptaddress,floatfix,a4paper,footinbib,bibnotes,prb]{revtex4}
\usepackage{amssymb}

\usepackage{amsmath}
\usepackage{graphicx}
\usepackage{dcolumn}
\usepackage{bm}

\begin{document}

\title{Electron-Phonon Correlation Effects in Molecular Transistors}

\author{C. A. Balseiro}
\affiliation{Instituto Balseiro and Centro At\'omico Bariloche, Comisi\'on
Nacional de Energ\'{\i}a At\'omica, 8400 Bariloche, Argentina.}
\author{P. S. Cornaglia}
\affiliation{Centre de Physique Th\'eorique, \'Ecole Polytechnique, CNRS, 91128 Palaiseau Cedex, France.}
\author{D. R. Grempel\footnote{Deceased}}
\affiliation{CEA-Saclay, DSM/DRECAM/SPCSI, B\^at. 462, F91191 Gif sur Yvette, France.}

\date{\today}

\begin{abstract}
The interplay of electron-electron and electron-phonon interactions is
studied analytically in the Kondo regime. A Holstein electron-phonon
coupling is shown to produce a weakening of the gate voltage dependence of
the Kondo temperature and may explain the observed anomalies in some of these
devices. 
A molecular center-of-mass mode opens a new channel for charge and spin
fluctuations and in the antiadabatic limit the latter are described by an
asymmetric two-channel Kondo model. Below the Kondo temperature the system
develops a dynamical Jahn-Teller distortion and a low energy peak emerges in
the phonon spectral density that could be observed in Raman microscopy
experiments.
\end{abstract}



\maketitle

\section{Introduction}

One of the most exciting aspects in the physics of molecular transistors is
the interplay between electron-electron and electron-phonon (\textit{e}-ph)
interactions.\cite{Nitzan2003} Molecules attached to leads act like small
quantum dots with broadened but well defined electronic levels and large
charging energies. Electron-electron interactions lead to the well known
Coulomb blockade effect \cite{JPark_2002,Kubatkin_2003} and the strongly
correlated Kondo physics. \cite{WLiang_2002,JPark_2002,Yu2004,Yu2005} When
the coupling of electronic degrees of freedom and molecular vibrational
modes is strong, a rich variety of behavior emerges.\cite
{JPark_2002,Kubatkin_2003,WHoXX,HPark_2000,WLiang_2002,Yu2004,Yu2005,%
Nitzan2003,Braig2003,Mitra_2004,Cornaglia2004,Cornaglia2005a,%
Cornaglia2005b,paaske:176801,arrachea:041301,mravlje2005,al-hassanieh2005,%
koch:206804,koch-2006-96}
The presence of steps in the current-voltage characteristics of molecular
transistors can be explained by the coupling to a well defined phononic
mode. Other features, like the anomalous behavior of the Kondo temperature
as a function of the gate voltage, remain yet to be understood.\cite
{WLiang_2002,Yu2005} Center-of-mass oscillations of the active component of
a transistor and electron-electron interactions are expected to produce a 
\textit{shuttling} effect in nanomechanical devices. \cite{gorelik:4526} The
interplay of such mode with strong correlations at low temperatures is quite
a complicated problem of general interest and experimental relevance.\cite
{HPark_2000}

\begin{figure}[tbp]
\includegraphics[width=7cm,clip=true]{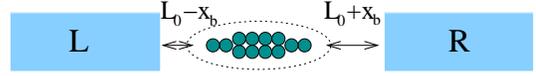}
\caption{Scheme of the molecular device studied in this work. A molecule
is attached to contacts: $L$ and $R$, the center-of-mass motion
modulates the molecule-leads tunneling barriers.} \label{fig:fig1}
\end{figure}

Here we consider a molecule in the Kondo regime with a Holstein coupling and
a center-of-mass motion mode (CMM) that modulates asymmetrically the
coupling between the molecule and the leads (see Fig. \ref{fig:fig1}). The
rest of the paper is organized as follows: in next section we present the
model and obtain the Kondo Hamiltonians describing the low energy physics,
in section III we first analyze the behavior of Kondo couplings and Kondo
temperatures, then we present variational wave functions that allow to
explore a wider range of parameters, in particular situations in which the
phonon frequencies are of the order or smaller than the Kondo temperature.
Section IV contains a study of the spectral density of the CMM phonon.
Finally, last section summarizes results and conclusions.

\section{Model}

The model Hamiltonian is $H=H_{M}+H_{E}+H_{ME}$ where the first two terms
describe the isolated molecule and electrodes, respectively, and the last
term describes their coupling. We have 
\begin{eqnarray}
H_{M} &=&\varepsilon _{d}n_{d}+Un_{d\uparrow }n_{d\downarrow }-\lambda
\left( n_{d}-1\right) \left( a^{{}}+a^{\dagger }\right) \\
&+&\omega _{0}a^{\dagger }a^{{}}+\omega _{1}b^{\dagger }b^{{}},  \nonumber \\
H_{E} &=&\sum_{{k},\sigma ,\alpha }\varepsilon _{\alpha {k}}\;c_{\alpha {k}%
\sigma }^{\dagger }c_{\alpha {k}\sigma }^{{}}, \\
H_{ME} &=&\sum_{{k},\sigma ,\alpha }\;V_{\alpha }(x_{b})\left( d_{\sigma
}^{\dagger }\;c_{\alpha {k}\sigma }^{{}}+c_{\alpha {k}\sigma }^{\dagger
}\;d_{\sigma }^{{}}\right) .
\end{eqnarray}
Here $n_{d}={n}_{d\uparrow }+{n}_{d\downarrow }$, ${n}_{d\sigma }=d_{\sigma
}^{\dagger }d_{\sigma }$, $d_{\sigma }^{\dagger }$ creates an electron at
the molecular orbital with energy $\varepsilon _{d}$, $U$ is the
intralevel Coulomb repulsion and $c_{\alpha k\sigma }^{\dagger }$
creates an electron in the $k$ mode of lead $\alpha =R$, $L$. The operators $%
a^{\dagger }$ and $b^{\dagger }$ create a phonon excitation on the Holstein
mode with energy $\omega _{0}$ and on the CMM with energy $\omega _{1}$,
respectively. The single particle energies are measured from the Fermi
energy and $\hbar =1$.

From hereon we consider a symmetric molecule with identical $R$ and $L$
leads ($\varepsilon_{\alpha k}\equiv \varepsilon_k$) and take $%
V_{R}(x_b)=V_{0}+g x_b$ and $V_{L}(x_b)=V_{0}-g x_b$, where $x_b=(1/2M\omega
_{1})^{1/2}(b+b^{\dagger})$ is the molecular center-of-mass coordinate. For
a molecule with inversion symmetry, $x_b$ cannot couple linearly with the
electron density.

It is convenient to rewrite the Hamiltonian in terms of symmetric and
antisymmetric modes defined by $c_{Sk\sigma }=(c_{Rk\sigma }+ c_{Lk\sigma })/%
\sqrt{2}$ and $c_{Ak\sigma }=(c_{Rk\sigma }- c_{Lk\sigma })/\sqrt{2}$: 
\begin{eqnarray}
H_{ME} &=&V_{S}\sum_{k,\sigma }(d_{\sigma }^{\dagger}c_{Sk\sigma
}+c_{Sk\sigma }^{\dagger}d_{\sigma })  \nonumber \\
&+&V_{A}(b+b^{\dagger})\sum_{k,\sigma }(d_{\sigma }^{\dagger}c_{Ak\sigma
}+c_{Ak\sigma }^{\dagger}d_{\sigma })  \label{Hhyb}
\end{eqnarray}
with $V_{S}=\sqrt{2}V_{0}$ and $V_{A}=\sqrt{1 /M\omega _{1}}g$. This shows
that the symmetric mode is directly coupled to the molecular orbital while
the antisymmetric coupling is phonon-assisted.

In the Kondo limit we eliminate the charge fluctuations by means of a
Schrieffer-Wolff transformation extending the analysis of Refs.~[%
\onlinecite{Cornaglia2004,Cornaglia2005a,Stephan1997}]. We obtain that the spin
fluctuations are described by an asymmetric two-channel Kondo Hamiltonian 
\begin{equation}
H_{K}=\sum_{\nu =A,S}J_{\nu }\sum_{k,k^{\prime },\sigma ,\sigma ^{\prime }}%
\vec{S}\cdot c_{\nu k\sigma }^{\dagger }\frac{{\vec{\sigma}_{\sigma \sigma
^{\prime }}}}{2}c_{\nu k^{\prime }\sigma ^{\prime }}.  \label{eq:2cKondo}
\end{equation}
Here $\vec{S}$ is the spin operator of the molecular orbital, $\vec{\sigma}$
are the Pauli matrices, and 
\begin{eqnarray}
J_{S} &=&\sum_{m=0}^{\infty }\left( \frac{2V_{S}^{2}\gamma _{m}}{-\widetilde{%
\varepsilon }_{d}+m\omega _{0}}+\frac{2V_{S}^{2}\gamma _{m}}{\widetilde{%
\varepsilon }_{d}+\widetilde{U}+m\omega _{0}}\right) ,  \nonumber
\label{eq:JS} \\
&& \\
J_{A} &=&\sum_{m=0}^{\infty }\left( \frac{2V_{A}^{2}\gamma _{m}}{-\widetilde{%
\varepsilon }_{d}+\omega _{1}+m\omega _{0}}+\frac{2V_{A}^{2}\gamma _{m}}{%
\widetilde{\varepsilon }_{d}+\widetilde{U}+\omega _{1}+m\omega _{0}}\right) ,
\nonumber
\end{eqnarray}
where $\gamma _{m}=e^{-(\lambda /\omega _{0})^{2}}(\lambda /\omega
_{0})^{2m}/m!$, $\widetilde{\varepsilon }_{d}=\varepsilon _{d}+\lambda
^{2}/\omega _{0}$, $\widetilde{U}=U-2\lambda ^{2}/\omega _{0}$, $\widetilde{%
\varepsilon }_{d}<0$, and $\widetilde{U}>0$. To second order in $H_{ME}$
there are also inelastic contributions proportional to $V_{S}V_{A}$ which
couple the ground state of the isolated molecule to a molecular state with a
CMM excitation. The energy level structure of the molecule is analogous to
that of a multilevel quantum dot with level splitting $\omega _{1}$, and for
large $\omega _{1}$ these contributions can be neglected.\cite{Boese2002}
We will present below a more detailed calculation where this assumption is
not required.

\section{Spin Screening in the Kondo Limit}

The low energy properties of $H_{K}$ are well known.\cite{Nozieres1980} At
low temperatures only one channel participates in the screening of the
molecular spin, and the Fermi liquid picture applies as $T\rightarrow 0$. In
the renormalization group language, the zero temperature fixed point
corresponds to a diverging ratio of the largest to the smallest Kondo
coupling. Only for $J_{S}=J_{A}$ both channels participate in an
overscreening of the molecular spin. In our context, this would be a quite
peculiar situation. In the moderate or weak \textit{e}-ph coupling regime $%
V_{A}$ is smaller than $V_{S}$ and the Kondo coupling $J_{S}$ is therefore
larger than $J_{A}$. The screening of the molecular spin is then dominated
by the symmetric channel. In this case the Kondo temperature is given by $%
T_{K}=De^{-1/\rho J_{S}}$, where $D$ is a high energy cutoff and 
$\rho $ is the bare electronic density of states of the leads at the Fermi
level. These results imply that, at zero temperature, the asymmetric channel
is effectively decoupled from the molecule and the conductance is governed
by the symmetric channel. As a consequence the conductance is simply $%
2e^{2}/h$.
\begin{figure}[tbp]
\includegraphics[width=8.5cm,clip=true]{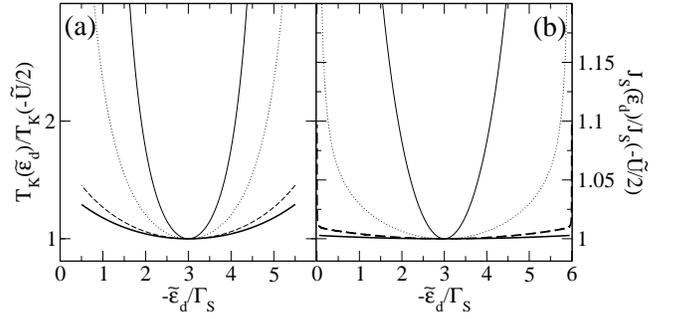}
\caption{a) The Kondo temperature plotted against $\widetilde{\varepsilon}_{d}$  for different values of the {\it e}-ph coupling
 $\lambda$. $\lambda/\omega_0=1$ (thin line), $2.5$ (dotted line), $3.5$ 
(dashed line), and $4.5$ (thick line). Here $\Gamma_S =\pi\rho V_S^2= 20meV$, 
$\widetilde{U}=6\Gamma_S$, and $\omega_0/\widetilde{U}=0.5$. b) Kondo coupling $J_S$ against $\widetilde{\varepsilon}_{d}$ for the parameters of a).
 } \label{fig:Tk_ed}
\end{figure}

In what follows we show that the obtained $J_{S}$ may explain some
experimentally observed anomalies on the gate voltage dependence of $T_{K}$
.\cite{WLiang_2002,Yu2005} In the absence of \textit{e}-ph coupling ($%
\lambda =0$) $J_{S}$ is given by the usual expression $J_{S}(\lambda=0)=-2V_{S}^{2}U/[%
\varepsilon _{d}(\varepsilon _{d}+U)]$ and increases as the gate voltage
decreases $\left| \varepsilon _{d}\right| $ or $(\varepsilon _{d}+U)$. The
experimentally observed dependence of the Kondo temperature on the gate
voltage is, however, much weaker than what is inferred from this expression.
In Ref.~\onlinecite{Yu2005} the Kondo temperature remains almost independent 
of the gate voltage and presents a fast increase very close to the charge
 degeneracy points. 
Other systems also present a change in the Kondo temperature dependence
on gate voltage at the borders of the Coulomb blockade valley.\cite{WLiang_2002}

These anomalies can be understood by the
presence of  \textit{e}-ph coupling. Using the
expression for $J_{S}$ from Eq.~(\ref{eq:JS}) we plotted the gate voltage
dependence of the Kondo temperature for different values of the \textit{e}%
-ph coupling (see Fig.~\ref{fig:Tk_ed}). The gate voltage dependence of $%
T_{K}$ weakens monotonically with increasing \textit{e}-ph interaction.
This can be seen more easily making some simplifying approximations on Eq.~(%
\ref{eq:JS}) for $\widetilde{\varepsilon }_{d}\sim -\widetilde{U}/2$. Note
that the relevant quantities here are $\widetilde{\varepsilon }_{d}$ and $%
\widetilde{U}$ since the valence changes in the molecule occur for $%
\widetilde{\varepsilon }_{d}=0$ and $\widetilde{\varepsilon }_{d}=-%
\widetilde{U}$. One of the effects of the  \textit{e}-ph
coupling is to decreases both $\left| \widetilde{\varepsilon }_{d}\right| \ $
and $\ \widetilde{U}$ shrinking the region of valence stability. For large $%
\lambda /\omega _{0}\gg 1$, $\gamma (m)\equiv \gamma _{m}$ is approximately
Gaussian peaked at $m\sim m^{\star }=(\lambda /\omega _{0})^{2}$ of width $%
\sqrt{m^{\star }}$ and satisfies $\sum_{m}\gamma _{m}=1$. In this case, the
series in Eq.~(\ref{eq:JS}) can be approximated with 
\begin{eqnarray}
J_{S}&\simeq&
2V_{S}^{2}\left(\frac{1}{-\widetilde{\varepsilon }_{d}+m^{\star }\omega _{0}}+\frac{1}{
\widetilde{U}+\widetilde{\varepsilon }_{d}+m^{\star }\omega _{0}}\right)\\
&\simeq& 2V_{S}^{2}\left(\frac{1}{-\varepsilon_d}+\frac{1}{U+\varepsilon_d}\right).
\end{eqnarray}
We obtain in the large $\lambda/\omega_0$ limit that $J_S$
is given by the $\lambda=0$ expression!
This result, however, 
ceases to be valid close to the charge degeneracy points $\widetilde{\varepsilon }_{d}\rightarrow 0,-%
\widetilde{U}$ where the $m=0$ term in Eq.~(\ref{eq:JS}) diverges with an
exponentially small Franck-Condon prefactor $\gamma_0=e^{-(\lambda
/\omega _{0})^{2}}.$ The smallness of this prefactor requires $\widetilde{%
\varepsilon }_{d}$ or $\widetilde{\varepsilon}_d+\widetilde{U}$ to be exponentially small in order to make the $m=0$ term of order one [see Fig.~\ref{fig:Tk_ed}(b)]. 
As a consequence, $J_S$ behaves as $J_S(\lambda=0)$ which is regular at 
the charge degeneracy points and only very close to them in behaves
as $J_S\sim-2V_{S}^{2}\widetilde{U}%
/[\widetilde{\varepsilon} _{d}(\widetilde{\varepsilon} _{d}+\widetilde{U})]$,
which is divergent.
The latter expression gives the behavior of $J_S$ that can naively
 expected using the renormalized single particle energy 
$\widetilde{\varepsilon}_d$ and Hubbard 
interaction $\widetilde{U}$ in the usual expression for the Kondo
 coupling in an Anderson model. 
The behavior of the Franck-Condon matrix elements $\gamma_m$ is 
the source of the anomalous dependence of the Kondo temperature
on the gate voltage. Related anomalies are obtained in the valence change
 of simpler non-interacting models \cite{Braig2003} and in the incoherent
 non-linear transport.\cite{koch:206804}

It is important to notice that to lowest order in the hybridization only the
Holstein coupling renormalizes $J_S$ and produces the anomalous behavior. 
In what follows we show that in higher
order the CMM contributes to increase the Kondo coupling an effect that
may help explain the high values of $T_K$ observed in some devices. To do this
we now return to the original description given by the full Hamiltonian $H$
and build the ground state wavefunction using a generalized Varma-Yafet
ansatz.\cite{Varma1976} While this method is exact only in the large $N$
limit, where the spin symmetry is extended to the $SU(N)$ group, it gives
the correct zero-temperature physics even in the spin $1/2$ ($N=2$) case.
Following Ref. \cite{Alascio1988} we propose for the ground state
wavefunction: 
\begin{equation}
\left| \Phi \right\rangle = b_{0}\left| \phi _{F}\right\rangle \left|
P_{0}\right\rangle +\sum_{k,\sigma,\nu=A,S } b_{\nu k}\left| \phi _{\nu
k}\right\rangle \left| P_{\nu k}\right\rangle,
\end{equation}
where $\left| \phi _{F}\right\rangle $ describes the Fermi sea with total
energy $E_{FS}$, $\left| \phi _{\nu k}\right\rangle =\frac{1}{\sqrt{2}}%
\sum_\sigma d_{\sigma }^{\dagger}c_{\nu k\sigma }\left| \phi
_{F}\right\rangle $, $\left| P_{0}\right\rangle $, and $\left| P_{\nu
k}\right\rangle $ are phonon states including both the Holstein coupling and
the CMM, and we considered the $U\rightarrow \infty $ limit. The
coefficients $b_{0}$ and $b_{\nu k}$ are chosen for the wavefunction to
satisfy the Schr\"odinger equation $H\left| \Phi \right\rangle =E_{GS}\left|
\Phi \right\rangle $.

A simple procedure allows to determine the form of the phonon states and
their coefficients. First we note that $\langle \phi _{F}\left| H\right|
\Phi \rangle =E_{GS}b_{0}\left| P_{0}\right\rangle $ and $\langle \phi _{\nu
k}\left| H\right| \Phi \rangle =E_{GS}b_{\nu k}\left| P_{\nu k}\right\rangle 
$. This generates a system of coupled equations for the phonon states that,
after eliminating $b_{\nu k}\left| P_{\nu k}\right\rangle $, can be cast in
the form 
\begin{eqnarray}
(\delta E+\varepsilon _{d})\left| P_{0}\right\rangle
&=&[H_{M}^{0}+2V_{S}^{2}\Gamma (\widehat{E})  \nonumber \\
&+&2V_{A}^{2}(b+b^{\dagger })\Gamma (\widehat{E})(b+b^{\dagger })]\left|
P_{0}\right\rangle .
\end{eqnarray}
Here $\delta E=E_{GS}-E_{FS}-\varepsilon _{d}$ is the energy gained due to
the hybridization, $\widehat{E}=-\delta E-\varepsilon _{d}+H_{M}^{1}$, $%
H_{M}^{\ell }$ is the projection of $H_{M}$ on the subspace with $\ell $
electrons in the molecular orbital, and 
\begin{equation}
\Gamma (z)=\sum_{k\leq k_{F}}\frac{1}{\varepsilon _{k}-z}=\frac{1}{2D}\ln
\left| \frac{z}{D+z}\right| ,
\end{equation}
where we considered for simplicity a square half-filled conduction band of
total width $2D$. We will study first the case with $\lambda =0$. In this
case the state $\left| P_{0}\right\rangle $ describes only the CMM degrees
of freedom and can be expanded in the base of eigenstates of the uncoupled
phonon $\left| P_{0}\right\rangle =\sum_{n}\alpha _{n}\left| n\right\rangle $%
. The following recursion relation is obtained for the coefficients: 
\begin{equation}
M_{nn}\alpha _{n}=M_{n,n+2}\alpha _{n+2}+M_{n,n-2}\alpha _{n-2},
\end{equation}
where
\begin{subequations}\label{eq:coefs}
\begin{eqnarray}
M_{n,n} &=&\delta E+\varepsilon _{d}- \omega_{1}n-2V_{S}^{2}\Gamma _{n} 
\nonumber \\
&-&2V_{A}^{2}[(n+1)\Gamma _{n+1}+n\Gamma _{n-1}],\\
M_{n,n+2} &=&2V_{A}^{2}\sqrt{(n+1)(n+2)}\Gamma _{n+1}, \\
M_{n,n-2} &=&2V_{A}^{2}\sqrt{n(n-1)}\Gamma _{n-1},
\end{eqnarray}
\end{subequations}
with $\Gamma _{n}=\Gamma (-\delta E+\omega _{1}n)$. The phonon state $\left|
P_{0}\right\rangle $ can only include even or odd $n$ states and therefore
has a well defined parity. We can find approximate solutions of these
equations by noting that in the Kondo limit $\delta E$ is small. We have 
\begin{equation}
\Gamma _{0}\simeq \frac{1}{2D}\ln \left| \frac{\delta E}{D}\right| \text{and 
}\Gamma _{n}\simeq \frac{1}{2D}\ln \left| \frac{\omega _{1}n}{D+\omega _{1}n}%
\right| \text{ for }n>0,
\end{equation}
that give for the even-$\left| P_{0}\right\rangle $ subspace 
\begin{equation}
(\delta E+\varepsilon _{d}-2V_{S}^{2}\Gamma _{0}-2V_{A}^{2}\Gamma
_{1}-\Sigma _{S})\alpha _{0}=0,  \label{eq:selfcons}
\end{equation}
where $\Sigma _{S}$ is a small correction that can be written as a continued
fraction after eliminating all coefficients $\alpha _{n}$ with $n>0$.
Equation (\ref{eq:selfcons}) gives a simple self-consistent condition for $%
\delta E$ and in the weak coupling regime the usual solution is obtained 
\begin{equation}
\delta E=-De^{-\frac{1}{\rho J_{S}^{\prime }}},  \label{eq:TkS}
\end{equation}
where $\rho =(2D)^{-1}$ and 
\begin{equation}
J_{S}^{\prime }=\frac{2V_{S}^{2}}{\left| \varepsilon _{d}\right|
+2V_{A}^{2}\Gamma _{1}+\Sigma _{S}}.  \label{eq:varmaJs}
\end{equation}
The energy gain $\delta E$ is associated to the formation of the Kondo
singlet and is of the order of the Kondo temperature. This result shows
that, in the even-$\left| P_{0}\right\rangle $ subspace, the molecular spin
is Kondo screened by the symmetric channel, and the antisymmetric one only
contributes to a renormalization of the molecular energy. This is the ground
state for $V_{S}\gtrsim $ $V_{A}$. In the opposite situation the ground
state belongs to the odd-$\left| P_{0}\right\rangle $ subspace, and the
antisymmetric channel screens the molecular spin with a Kondo coupling
constant 
\begin{equation}
J_{A}^{\prime }=\frac{2V_{A}^{2}}{\left| \varepsilon _{d}\right|
+2V_{S}^{2}\Gamma _{1}+2V_{A}^{2}\Gamma _{2}+\Sigma _{A}+\omega _{1}}.
\label{eq:varmaJa}
\end{equation}
Except for the presence of molecular energy correction terms in the
denominators, $J_{S}^{\prime }$ and $J_{A}^{\prime }$ reproduce the Kondo
couplings of Eq.~(\ref{eq:JS}) particularized for $\lambda =0$ and $U\to
\infty $. The same procedure can be followed when both \textit{e}-ph
couplings are non-zero and the result of Eq.~(\ref{eq:JS}) recovered. This
analysis not only validates the results of Eq.~(\ref{eq:2cKondo}) and Eq.~(%
\ref{eq:JS}) for $\omega _{0},\omega _{1}\gg T_{K}$ but it also gives some
insight on the physics when $\omega _{0}$ or $\omega _{1}$ are of the order
or smaller than $T_{K}$. To illustrate this we consider $\lambda =0$ and $%
\omega _{1}\gtrsim T_{K}$. The dependence of $\Gamma _{1}$ on $\delta E$
[see Eqs. (\ref{eq:coefs})] now needs to be considered and the resulting
equation for $\delta E$ can be solved numerically. The main result is an
increasing $|\delta E|$ (an increasing Kondo temperature) with decreasing $%
\omega _{1}$, an effect that may be related to the high Kondo temperatures
observed in some devices.\cite{Yu2005}

\section{Phonon Spectral Density}

Having characterized the nature of the Kondo screening, we now analyze the
molecular vibrational modes. We define the phonon propagators in terms of
the phonon fields $\ \phi _{0}=a+a^{\dagger }$ and $\ \phi _{1}=b+b^{\dagger
}$ 
\begin{equation}
\mathcal{D}_{\eta }(t)\equiv \langle \langle \phi _{\eta }(t),\phi _{\eta
}(0)\rangle \rangle =-i\theta (t)\left\langle [\phi _{\eta }(t),\phi _{\eta
}(0)]\right\rangle
\end{equation}
for $\eta =0,1$. Calculating the self-energy to second order in the \textit{e%
}-ph interaction we obtain 
\begin{equation}
\mathcal{D_{\eta }}(\omega )=\frac{2\omega _{\eta }}{\omega ^{2}-\omega
_{\eta }^{2}-2\omega _{\eta }\Pi _{\eta }(\omega )},
\end{equation}
where $\mathcal{D_{\eta }}(\omega )$ indicates the Fourier transform of $%
\mathcal{D_{\eta }}(t)$. This quantity has been studied by Hewson \textit{et
al.} \cite{Hewson_2002} for a Holstein coupling, where the self-energy $\Pi
_{0}(\omega )$ is given by the charge susceptibility. In the Kondo limit
(for $\widetilde{U}>0$) it has little structure at low frequency and the
vibrational mode just acquires a small width due to the coupling with the
electrons at the molecular orbital. The self-energy for the CMM is given by 
\begin{eqnarray}
\Pi _{1}(\omega ) &=&2V_{A}^{2}\sum_{k,\sigma }(\langle \langle d_{\sigma
}^{\dagger }c_{Ak\sigma },c_{Ak\sigma }^{\dagger }d_{\sigma }\rangle \rangle
_{\omega }  \nonumber \\
&+&\langle \langle c_{Ak\sigma }^{\dagger }d_{\sigma },d_{\sigma }^{\dagger
}c_{Ak\sigma }\rangle \rangle _{\omega }).
\end{eqnarray}
\begin{figure}[htbp]
\includegraphics[width=8.5cm,clip=true]{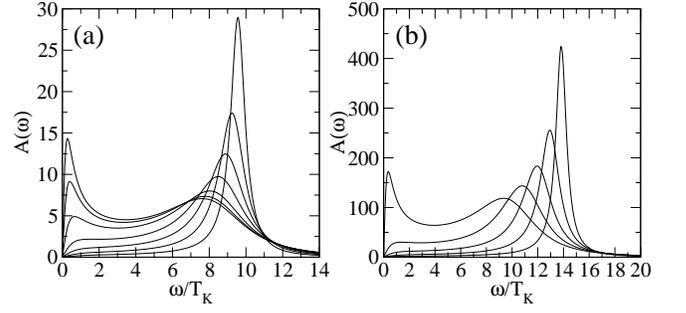}
\caption{ Phonon spectral density ${\cal A}(\omega)$. a) $\omega_1=0.25D$, $\omega_{1}/T_{K}=
10$, and $\varsigma /T_{K}= 0.15, 0.25, 0.35, 0.45, 0.55, 0.6$, and $0.625$ (from 
bottom to top in the low energy region). b) $\omega_1 = 0.025D$, $\omega_{1}/T_{K} = $
15,  $\varsigma /T_{K}=0.15, 0.25, 0.35, 0.45$, and $0.55$ (from 
bottom to top in the low energy region).} \label{fig:fig3}
\end{figure}
Since the Kondo screening is due to the symmetric channel, the antisymmetric
electronic modes can be decoupled from the molecular orbital and $\Pi
_{1}(\omega )$ calculated in the lowest order in $V_{A}$. $\Pi _{1}(\omega )$
can then be written in terms of one-particle propagators that for $T\ll
T_{K} $ and $\omega \ll \left| \varepsilon _{d}\right| $ are given by $-%
\frac{1}{\pi }\sum_{k}\mathrm{Im}\langle \langle c_{Ak\sigma },c_{Ak\sigma
}^{\dagger }\rangle \rangle _{\omega +i0}=\rho $ and 
\begin{equation}
-\frac{1}{\pi }\mathrm{Im}\langle \langle d_{\sigma },d_{\sigma }^{\dagger
}\rangle \rangle _{\omega +i0}=\frac{T_{K}}{\Gamma _{S}}\frac{T_{K}/\pi }{%
\omega ^{2}+T_{K}^{2}}.  \label{eq:Kpeak}
\end{equation}
We have
\begin{eqnarray*}
\Pi_1(\omega ) &=&2V_{A}^{2}\rho \frac{T_{K}}{\Delta }\int d\nu _{1}\int
d\nu _{2}\left[\frac{f(\nu _{1})-f(\nu _{2})}{\omega +\nu _{1}-\nu
_{2}+i0}\right. \\
&+&\left.\frac{f(\nu _{2})-f(\nu _{1})}{\omega -\nu _{1}+\nu _{2}+i0}\right]%
\frac{T_{K}/\pi }{\nu _{1}^{2}+T_{K}^{2}}.
\end{eqnarray*}
At zero temperature the integrals reduce to a simple form and the phonon
propagator reads:
\[
\mathcal{D}_{1}(\omega )=\frac{2\omega _{1}}{\omega ^{2}-\omega
_{1}^{2}-2\omega _{1}\varsigma [\ln (\frac{\omega ^{2}+T_{K}^{2}}{D^{2}}%
)-i2\arctan (\frac{\omega }{T_{K}})]} 
\]
with $\varsigma =2V_{A}^{2}\rho T_{K}/\Gamma
_{S}=(V_{A}/V_{S})^{2}2T_{K}/\pi $.

The phonon spectral density $\mathcal{A}(\omega )=-\mathop{\rm Im}{\cal D}%
_{1}(\omega )/\pi $ is shown in Fig. \ref{fig:fig3} for different values of
the microscopic parameters. For small \textit{e}-ph coupling it presents
a single peak at $\sim \omega_1$ broadened by the coupling to the electronic
degrees of freedom. For temperatures lower than $T_{K}$ and moderate 
\textit{e}-ph couplings, the system presents a soft mode (low energy peak). 
For temperatures larger than $T_{K}$ (not shown) the soft mode disappears as the
low-frequency spectral density of the $d$-electrons is structureless and
small.

\begin{figure}[htbp]
\includegraphics[width=8.5cm,clip=true]{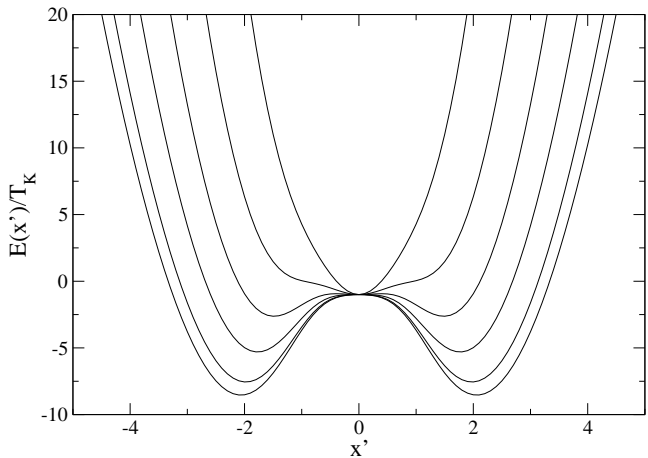}
\caption{ Energy gain as a function of the reduced displacement $x^\prime=\frac{g}{V_0}x_b$
of the CMM for different values of the electron phonon coupling 
$\varsigma /T_{K}= 0.15, 0.25, 0.35, 0.45, 0.55, 0.6$, and $0.625$ (from top to bottom).
Other parameters as in Fig.~\ref{fig:fig3}a.} \label{fig:softmode}
\end{figure}

The emergence of the soft mode is associated to the onset of
 a dynamical Jahn-Teller
distortion \cite{Alascio1988} induced by Kondo correlations.
To illustrate this we perform a semiclassical treatment in which
the CMM displacement $x_b$ is replaced by a real number. For 
simplicity we will consider only the case $\lambda=0$, but 
the conclusions are valid in the general case.
The molecule is coupled to the left and right leads with 
hopping amplitudes $V_L=(V_0+gx_b)$ and $V_R=(V_0-gx_b)$, respectively.
For a given value of the C-number $x_b$ we can perform a unitary transformation
\begin{equation}
\left(
\begin{array}{c}
 c_{1k\sigma} \\  c_{2k\sigma} \end{array}\right)=
\frac{1}{\sqrt{V_L^2+V_R^2}}
\left( \begin{array}{cc}
 V_L& V_R \\ -V_R & V_L \end{array}\right)
\left(\begin{array}{c}
 c_{Lk\sigma} \\  c_{Rk\sigma} \end{array}\right).
\end{equation}
to a new base where the $\{c_{2k\sigma}\}$ are decoupled from the
 molecule and play no role
in the Kondo physics while the $\{c_{1k\sigma}\}$
are coupled with a tunneling amplitude $\sqrt{2}V_0\sqrt{1+\frac{g^2}{V_0^2}x_b^2}$
that determines the Kondo screening. The Kondo temperature for this
semiclassical treatment is therefore given by
\begin{equation}
T_K^{C}=De^{-1/[\rho J_S(1+\frac{g^2}{V_0^2}x_b^2)]}.
\end{equation}
We can write the zero-temperaure energy gain of the system as a sum of the
 electronic [see Eq.~(\ref{eq:TkS})] and phononic contributions\cite{Alascio1988,clougherty:035507}
\begin{eqnarray}\label{eq:engain}
E(x^\prime) &=& -De^{-1/[\rho J_S (1+{x^\prime}^2)]}+\frac{1}{2 \pi} \omega_1 \frac{T_K}{\varsigma} {x^\prime}^2,
\end{eqnarray}
where we have introduced the reduced displacement $x^\prime =\frac{g}{V_0}x_b$.
In Fig.~\ref{fig:softmode} we plot the energy gain as a function of $x^\prime$
for the same parameters as in Fig.~\ref{fig:fig3}a. For small \textit{e}-ph
coupling, $\varsigma/T_K\ll 1$, the energy has a parabolic form and the minimum 
is for $x^\prime=0$.
The curvature of the parabola decreases with increasing $\varsigma$
and this is consistent with the shift to lower energies of the main peak in the spectral
density (see Fig.~\ref{fig:fig3}a). For $\varsigma/T_K \gtrsim 0.3$ two symmetric minima emerge 
for non-zero values of $x^\prime$ showing that there is there is
a Jahn-Teller instability associated with the formation of the Kondo effect.
This Jahn-Teller distortion has associated a soft mode which is reflected in the
phonon spectral density as a low energy peak.

\section{Summary and Conclusions}

We have studied analytically electron--phonon correlation effects in
molecular transistors using Schrieffer-Wolff transformations and a
Varma-Yafet ansatz. While the first technique provides a description in
terms of low energy Hamiltonians, the latter allows the study of the whole
range of parameters, including situations in which the phonon frequencies
are smaller than the Kondo temperature. We considered a Holstein coupling
and a center-of-mass mode and showed that both produce a renormalization of the
Kondo temperature. Remarkably, the Holstein coupling renormalizes the Kondo
coupling and weakens the gate voltage dependence of $T_{K}$. This may
explain the puzzling behavior observed in molecular Co and Cu complexes.
\cite{WLiang_2002,Yu2005}

The center-of-mass motion of the molecule opens an additional channel for
charge and spin fluctuations. At low energies the latter are described by an
asymmetric two-channel Kondo Hamiltonian. 
Due to channel asymmetry, however, the screening of the localized spin will
be in general dominated by a single channel. 
While the temperature dependence of the conductance may present some
anomalies due to the additional channel, at low temperatures the system is a
Fermi liquid and the low bias conductance through the molecule is $G\sim
2e^2/h$. Our calculations show that, in the Kondo regime a single channel is
active at zero temperature and no interference is possible between direct
and phonon assisted channels~\endnote{Such interference was proposed in
Ref.~[\onlinecite{al-hassanieh2005}] where the authors considered a single
phonon mode that modulates \textit{both} the molecular level and the
molecule-leads coupling.}.

Finally, we showed that the phonon spectral density of the center-of-mass
mode presents a low energy structure induced by the coupling to the Kondo
quasiparticles. Recent technical advances in single molecule phonon
spectroscopy \cite{hayazawa2002,hartschuh2003} may allow for a direct study
of this effect and provide useful information on the \textit{e}-ph
interaction in these systems.


\begin{thebibliography}{29}
\expandafter\ifx\csname natexlab\endcsname\relax\def\natexlab#1{#1}\fi
\expandafter\ifx\csname bibnamefont\endcsname\relax
  \def\bibnamefont#1{#1}\fi
\expandafter\ifx\csname bibfnamefont\endcsname\relax
  \def\bibfnamefont#1{#1}\fi
\expandafter\ifx\csname citenamefont\endcsname\relax
  \def\citenamefont#1{#1}\fi
\expandafter\ifx\csname url\endcsname\relax
  \def\url#1{\texttt{#1}}\fi
\expandafter\ifx\csname urlprefix\endcsname\relax\def\urlprefix{URL }\fi
\providecommand{\bibinfo}[2]{#2}
\providecommand{\eprint}[2][]{\url{#2}}

\bibitem[{\citenamefont{Nitzan and Ratner}(2003)}]{Nitzan2003}
\bibinfo{author}{\bibfnamefont{A.}~\bibnamefont{Nitzan}} \bibnamefont{and}
  \bibinfo{author}{\bibfnamefont{M.~A.} \bibnamefont{Ratner}},
  \bibinfo{journal}{Science} \textbf{\bibinfo{volume}{300}},
  \bibinfo{pages}{1384} (\bibinfo{year}{2003}).

\bibitem[{\citenamefont{Park et~al.}(2002)\citenamefont{Park, Pasupathy,
  Goldsmith, Chang, Yaish, Petta, Rinkoski, Sethna, Abru{\~n}a, McEuen
  et~al.}}]{JPark_2002}
\bibinfo{author}{\bibfnamefont{J.}~\bibnamefont{Park}},
  \bibinfo{author}{\bibfnamefont{A.~N.} \bibnamefont{Pasupathy}},
  \bibinfo{author}{\bibfnamefont{J.~I.} \bibnamefont{Goldsmith}},
  \bibinfo{author}{\bibfnamefont{C.}~\bibnamefont{Chang}},
  \bibinfo{author}{\bibfnamefont{Y.}~\bibnamefont{Yaish}},
  \bibinfo{author}{\bibfnamefont{J.~R.} \bibnamefont{Petta}},
  \bibinfo{author}{\bibfnamefont{M.}~\bibnamefont{Rinkoski}},
  \bibinfo{author}{\bibfnamefont{J.~P.} \bibnamefont{Sethna}},
  \bibinfo{author}{\bibfnamefont{H.~D.} \bibnamefont{Abru{\~n}a}},
  \bibinfo{author}{\bibfnamefont{P.~L.} \bibnamefont{McEuen}},
  \bibnamefont{et~al.}, \bibinfo{journal}{Nature}
  \textbf{\bibinfo{volume}{417}}, \bibinfo{pages}{722} (\bibinfo{year}{2002}).

\bibitem[{\citenamefont{Kubatkin et~al.}(2003)\citenamefont{Kubatkin, Danilov,
  Hjort, Cornil, Br{\'e}das, Stuhr-Hansen, Hedeg{\aa}rd, and
  Bj{\o}rnholm}}]{Kubatkin_2003}
\bibinfo{author}{\bibfnamefont{S.}~\bibnamefont{Kubatkin}},
  \bibinfo{author}{\bibfnamefont{A.}~\bibnamefont{Danilov}},
  \bibinfo{author}{\bibfnamefont{M.}~\bibnamefont{Hjort}},
  \bibinfo{author}{\bibfnamefont{J.}~\bibnamefont{Cornil}},
  \bibinfo{author}{\bibfnamefont{J.-L.} \bibnamefont{Br{\'e}das}},
  \bibinfo{author}{\bibfnamefont{N.}~\bibnamefont{Stuhr-Hansen}},
  \bibinfo{author}{\bibfnamefont{P.}~\bibnamefont{Hedeg{\aa}rd}},
  \bibnamefont{and}
  \bibinfo{author}{\bibfnamefont{T.}~\bibnamefont{Bj{\o}rnholm}},
  \bibinfo{journal}{Nature} \textbf{\bibinfo{volume}{425}},
  \bibinfo{pages}{698} (\bibinfo{year}{2003}).

\bibitem[{\citenamefont{Liang et~al.}(2002)\citenamefont{Liang, Shores,
  Bockrath, Long, and Park}}]{WLiang_2002}
\bibinfo{author}{\bibfnamefont{W.}~\bibnamefont{Liang}},
  \bibinfo{author}{\bibfnamefont{M.~P.} \bibnamefont{Shores}},
  \bibinfo{author}{\bibfnamefont{M.}~\bibnamefont{Bockrath}},
  \bibinfo{author}{\bibfnamefont{J.~R.} \bibnamefont{Long}}, \bibnamefont{and}
  \bibinfo{author}{\bibfnamefont{H.}~\bibnamefont{Park}},
  \bibinfo{journal}{Nature} \textbf{\bibinfo{volume}{417}},
  \bibinfo{pages}{725} (\bibinfo{year}{2002}).

\bibitem[{\citenamefont{Yu and Natelson}(2004)}]{Yu2004}
\bibinfo{author}{\bibfnamefont{L.~H.} \bibnamefont{Yu}} \bibnamefont{and}
  \bibinfo{author}{\bibfnamefont{D.}~\bibnamefont{Natelson}},
  \bibinfo{journal}{Nano Letters} \textbf{\bibinfo{volume}{4(1)}},
  \bibinfo{pages}{79} (\bibinfo{year}{2004}).

\bibitem[{\citenamefont{Yu et~al.}(2005)\citenamefont{Yu, Keane, Ciszek, Cheng,
  Tour, Baruah, Pederson, and Natelson}}]{Yu2005}
\bibinfo{author}{\bibfnamefont{L.~H.} \bibnamefont{Yu}},
  \bibinfo{author}{\bibfnamefont{Z.~K.} \bibnamefont{Keane}},
  \bibinfo{author}{\bibfnamefont{J.~W.} \bibnamefont{Ciszek}},
  \bibinfo{author}{\bibfnamefont{L.}~\bibnamefont{Cheng}},
  \bibinfo{author}{\bibfnamefont{J.~M.} \bibnamefont{Tour}},
  \bibinfo{author}{\bibfnamefont{T.}~\bibnamefont{Baruah}},
  \bibinfo{author}{\bibfnamefont{M.~R.} \bibnamefont{Pederson}},
  \bibnamefont{and} \bibinfo{author}{\bibfnamefont{D.}~\bibnamefont{Natelson}},
  \bibinfo{journal}{Phys. Rev. Lett.} \textbf{\bibinfo{volume}{95}},
  \bibinfo{pages}{256803} (\bibinfo{year}{2005}).

\bibitem[{\citenamefont{Ho}(2002)}]{WHoXX}
\bibinfo{author}{\bibfnamefont{W.}~\bibnamefont{Ho}}, \bibinfo{journal}{J.
  Chem. Phys.} \textbf{\bibinfo{volume}{55}}, \bibinfo{pages}{11033}
  (\bibinfo{year}{2002}).

\bibitem[{\citenamefont{Park et~al.}(2000)\citenamefont{Park, Park, Lim,
  Anderson, Alivisatos, and McEuen}}]{HPark_2000}
\bibinfo{author}{\bibfnamefont{H.}~\bibnamefont{Park}},
  \bibinfo{author}{\bibfnamefont{J.}~\bibnamefont{Park}},
  \bibinfo{author}{\bibfnamefont{A.~K.~L.} \bibnamefont{Lim}},
  \bibinfo{author}{\bibfnamefont{E.~H.} \bibnamefont{Anderson}},
  \bibinfo{author}{\bibfnamefont{A.~P.} \bibnamefont{Alivisatos}},
  \bibnamefont{and} \bibinfo{author}{\bibfnamefont{P.~L.}
  \bibnamefont{McEuen}}, \bibinfo{journal}{Nature}
  \textbf{\bibinfo{volume}{407}}, \bibinfo{pages}{57} (\bibinfo{year}{2000}).

\bibitem[{\citenamefont{Braig and Flensberg}(2003)}]{Braig2003}
\bibinfo{author}{\bibfnamefont{S.}~\bibnamefont{Braig}} \bibnamefont{and}
  \bibinfo{author}{\bibfnamefont{K.}~\bibnamefont{Flensberg}},
  \bibinfo{journal}{Phys. Rev. B} \textbf{\bibinfo{volume}{68}},
  \bibinfo{pages}{205324} (\bibinfo{year}{2003}).

\bibitem[{\citenamefont{Mitra et~al.}(2004)\citenamefont{Mitra, Aleiner, and
  Millis}}]{Mitra_2004}
\bibinfo{author}{\bibfnamefont{A.}~\bibnamefont{Mitra}},
  \bibinfo{author}{\bibfnamefont{I.}~\bibnamefont{Aleiner}}, \bibnamefont{and}
  \bibinfo{author}{\bibfnamefont{A.~J.} \bibnamefont{Millis}},
  \bibinfo{journal}{Phys. Rev. B} \textbf{\bibinfo{volume}{69}},
  \bibinfo{pages}{245302} (\bibinfo{year}{2004}).

\bibitem[{\citenamefont{Cornaglia et~al.}(2004)\citenamefont{Cornaglia, Ness,
  and Grempel}}]{Cornaglia2004}
\bibinfo{author}{\bibfnamefont{P.~S.} \bibnamefont{Cornaglia}},
  \bibinfo{author}{\bibfnamefont{H.}~\bibnamefont{Ness}}, \bibnamefont{and}
  \bibinfo{author}{\bibfnamefont{D.~R.} \bibnamefont{Grempel}},
  \bibinfo{journal}{Phys. Rev. Lett.} \textbf{\bibinfo{volume}{93}},
  \bibinfo{pages}{147201} (\bibinfo{year}{2004}).

\bibitem[{\citenamefont{Cornaglia et~al.}(2005)\citenamefont{Cornaglia,
  Grempel, and Ness}}]{Cornaglia2005a}
\bibinfo{author}{\bibfnamefont{P.~S.} \bibnamefont{Cornaglia}},
  \bibinfo{author}{\bibfnamefont{D.~R.} \bibnamefont{Grempel}},
  \bibnamefont{and} \bibinfo{author}{\bibfnamefont{H.}~\bibnamefont{Ness}},
  \bibinfo{journal}{Phys. Rev. B} \textbf{\bibinfo{volume}{71}},
  \bibinfo{pages}{075320} (\bibinfo{year}{2005}).

\bibitem[{\citenamefont{Cornaglia and Grempel}(2005)}]{Cornaglia2005b}
\bibinfo{author}{\bibfnamefont{P.~S.} \bibnamefont{Cornaglia}}
  \bibnamefont{and} \bibinfo{author}{\bibfnamefont{D.~R.}
  \bibnamefont{Grempel}}, \bibinfo{journal}{Phys. Rev. B}
  \textbf{\bibinfo{volume}{71}}, \bibinfo{pages}{245326}
  (\bibinfo{year}{2005}).

\bibitem[{\citenamefont{Paaske and Flensberg}(2005)}]{paaske:176801}
\bibinfo{author}{\bibfnamefont{J.}~\bibnamefont{Paaske}} \bibnamefont{and}
  \bibinfo{author}{\bibfnamefont{K.}~\bibnamefont{Flensberg}},
  \bibinfo{journal}{Phys. Rev. Lett.} \textbf{\bibinfo{volume}{94}},
  \bibinfo{eid}{176801} (\bibinfo{year}{2005}).

\bibitem[{\citenamefont{Arrachea and Rozenberg}(2005)}]{arrachea:041301}
\bibinfo{author}{\bibfnamefont{L.}~\bibnamefont{Arrachea}} \bibnamefont{and}
  \bibinfo{author}{\bibfnamefont{M.~J.} \bibnamefont{Rozenberg}},
  \bibinfo{journal}{Phys. Rev. B} \textbf{\bibinfo{volume}{72}},
  \bibinfo{pages}{041301(R)} (\bibinfo{year}{2005}).

\bibitem[{\citenamefont{Mravlje et~al.}(2005)\citenamefont{Mravlje, Ramsak, and
  Rejec}}]{mravlje2005}
\bibinfo{author}{\bibfnamefont{J.}~\bibnamefont{Mravlje}},
  \bibinfo{author}{\bibfnamefont{A.}~\bibnamefont{Ramsak}}, \bibnamefont{and}
  \bibinfo{author}{\bibfnamefont{T.}~\bibnamefont{Rejec}},
  \bibinfo{journal}{Phys. Rev. B} \textbf{\bibinfo{volume}{72}},
  \bibinfo{pages}{121403(R)} (\bibinfo{year}{2005}).

\bibitem[{\citenamefont{Al-Hassanieh et~al.}(2005)\citenamefont{Al-Hassanieh,
  Busser, Martins, and Dagotto}}]{al-hassanieh2005}
\bibinfo{author}{\bibfnamefont{K.~A.} \bibnamefont{Al-Hassanieh}},
  \bibinfo{author}{\bibfnamefont{C.~A.} \bibnamefont{Busser}},
  \bibinfo{author}{\bibfnamefont{G.~B.} \bibnamefont{Martins}},
  \bibnamefont{and} \bibinfo{author}{\bibfnamefont{E.}~\bibnamefont{Dagotto}},
  \bibinfo{journal}{Phys. Rev. Lett.} \textbf{\bibinfo{volume}{95}},
  \bibinfo{pages}{256807} (\bibinfo{year}{2005}).

\bibitem[{\citenamefont{Koch and von Oppen}(2005)}]{koch:206804}
\bibinfo{author}{\bibfnamefont{J.}~\bibnamefont{Koch}} \bibnamefont{and}
  \bibinfo{author}{\bibfnamefont{F.}~\bibnamefont{von Oppen}},
  \bibinfo{journal}{Phys. Rev. Lett.} \textbf{\bibinfo{volume}{94}},
  \bibinfo{eid}{206804} (\bibinfo{year}{2005}).

\bibitem[{\citenamefont{Koch et~al.}(2006)\citenamefont{Koch, Raikh, and von
  Oppen}}]{koch-2006-96}
\bibinfo{author}{\bibfnamefont{J.}~\bibnamefont{Koch}},
  \bibinfo{author}{\bibfnamefont{M.~E.} \bibnamefont{Raikh}}, \bibnamefont{and}
  \bibinfo{author}{\bibfnamefont{F.}~\bibnamefont{von Oppen}},
  \bibinfo{journal}{Phys. Rev. Lett.} \textbf{\bibinfo{volume}{96}},
  \bibinfo{pages}{056803} (\bibinfo{year}{2006}).

\bibitem[{\citenamefont{Gorelik et~al.}(1998)\citenamefont{Gorelik, Isacsson,
  Voinova, Kasemo, Shekhter, and Jonson}}]{gorelik:4526}
\bibinfo{author}{\bibfnamefont{L.~Y.} \bibnamefont{Gorelik}},
  \bibinfo{author}{\bibfnamefont{A.}~\bibnamefont{Isacsson}},
  \bibinfo{author}{\bibfnamefont{M.~V.} \bibnamefont{Voinova}},
  \bibinfo{author}{\bibfnamefont{B.}~\bibnamefont{Kasemo}},
  \bibinfo{author}{\bibfnamefont{R.~I.} \bibnamefont{Shekhter}},
  \bibnamefont{and} \bibinfo{author}{\bibfnamefont{M.}~\bibnamefont{Jonson}},
  \bibinfo{journal}{Phys. Rev. Lett.} \textbf{\bibinfo{volume}{80}},
  \bibinfo{pages}{4526} (\bibinfo{year}{1998}).

\bibitem[{\citenamefont{Stephan et~al.}(1997)\citenamefont{Stephan, Capone,
  Grilli, and Castellani}}]{Stephan1997}
\bibinfo{author}{\bibfnamefont{W.}~\bibnamefont{Stephan}},
  \bibinfo{author}{\bibfnamefont{M.}~\bibnamefont{Capone}},
  \bibinfo{author}{\bibfnamefont{M.}~\bibnamefont{Grilli}}, \bibnamefont{and}
  \bibinfo{author}{\bibfnamefont{C.}~\bibnamefont{Castellani}},
  \bibinfo{journal}{Phys. Lett. A} \textbf{\bibinfo{volume}{227}},
  \bibinfo{pages}{120} (\bibinfo{year}{1997}).

\bibitem[{\citenamefont{Boese et~al.}(2002)\citenamefont{Boese, Hofstetter, and
  Schoeller}}]{Boese2002}
\bibinfo{author}{\bibfnamefont{D.}~\bibnamefont{Boese}},
  \bibinfo{author}{\bibfnamefont{W.}~\bibnamefont{Hofstetter}},
  \bibnamefont{and}
  \bibinfo{author}{\bibfnamefont{H.}~\bibnamefont{Schoeller}},
  \bibinfo{journal}{Phys. Rev. B} \textbf{\bibinfo{volume}{66}},
  \bibinfo{pages}{125315} (\bibinfo{year}{2002}).

\bibitem[{\citenamefont{Nozi{\`e}res and Blandin}(1980)}]{Nozieres1980}
\bibinfo{author}{\bibfnamefont{P.}~\bibnamefont{Nozi{\`e}res}}
  \bibnamefont{and} \bibinfo{author}{\bibfnamefont{A.}~\bibnamefont{Blandin}},
  \bibinfo{journal}{J. Physique} \textbf{\bibinfo{volume}{41}},
  \bibinfo{pages}{193} (\bibinfo{year}{1980}).

\bibitem[{\citenamefont{Varma and Yafet}(1976)}]{Varma1976}
\bibinfo{author}{\bibfnamefont{C.~M.} \bibnamefont{Varma}} \bibnamefont{and}
  \bibinfo{author}{\bibfnamefont{Y.}~\bibnamefont{Yafet}},
  \bibinfo{journal}{Phys. Rev. B} \textbf{\bibinfo{volume}{13}},
  \bibinfo{pages}{2950} (\bibinfo{year}{1976}).

\bibitem[{\citenamefont{Alascio et~al.}(1988)\citenamefont{Alascio, Balseiro,
  Ort{\'{\i}}z, Kiwi, and Lagos}}]{Alascio1988}
\bibinfo{author}{\bibfnamefont{B.}~\bibnamefont{Alascio}},
  \bibinfo{author}{\bibfnamefont{C.}~\bibnamefont{Balseiro}},
  \bibinfo{author}{\bibfnamefont{G.}~\bibnamefont{Ort{\'{\i}}z}},
  \bibinfo{author}{\bibfnamefont{M.}~\bibnamefont{Kiwi}}, \bibnamefont{and}
  \bibinfo{author}{\bibfnamefont{M.}~\bibnamefont{Lagos}},
  \bibinfo{journal}{Phys. Rev. B} \textbf{\bibinfo{volume}{38}},
  \bibinfo{pages}{4698} (\bibinfo{year}{1988}).

\bibitem[{\citenamefont{Hewson and Meyer}(2002)}]{Hewson_2002}
\bibinfo{author}{\bibfnamefont{A.~C.} \bibnamefont{Hewson}} \bibnamefont{and}
  \bibinfo{author}{\bibfnamefont{D.}~\bibnamefont{Meyer}}, \bibinfo{journal}{J.
  Phys.:Condens. Matter} \textbf{\bibinfo{volume}{14}}, \bibinfo{pages}{427}
  (\bibinfo{year}{2002}).

\bibitem[{\citenamefont{Clougherty}(2003)}]{clougherty:035507}
\bibinfo{author}{\bibfnamefont{D.~P.} \bibnamefont{Clougherty}},
  \bibinfo{journal}{Phys. Rev. Lett.} \textbf{\bibinfo{volume}{90}},
  \bibinfo{pages}{035507} (\bibinfo{year}{2003}).

\bibitem[{\citenamefont{Hayazawa et~al.}(2002)\citenamefont{Hayazawa, Tarun,
  Inouye, and Kawata}}]{hayazawa2002}
\bibinfo{author}{\bibfnamefont{N.}~\bibnamefont{Hayazawa}},
  \bibinfo{author}{\bibfnamefont{A.}~\bibnamefont{Tarun}},
  \bibinfo{author}{\bibfnamefont{Y.}~\bibnamefont{Inouye}}, \bibnamefont{and}
  \bibinfo{author}{\bibfnamefont{S.}~\bibnamefont{Kawata}},
  \bibinfo{journal}{J. Appl. Phys.} \textbf{\bibinfo{volume}{92}},
  \bibinfo{pages}{6983} (\bibinfo{year}{2002}).

\bibitem[{\citenamefont{Hartschuh et~al.}(2003)\citenamefont{Hartschuh,
  Sanchez, Xie, and Novotny}}]{hartschuh2003}
\bibinfo{author}{\bibfnamefont{A.}~\bibnamefont{Hartschuh}},
  \bibinfo{author}{\bibfnamefont{E.~J.} \bibnamefont{Sanchez}},
  \bibinfo{author}{\bibfnamefont{X.~S.} \bibnamefont{Xie}}, \bibnamefont{and}
  \bibinfo{author}{\bibfnamefont{L.}~\bibnamefont{Novotny}},
  \bibinfo{journal}{Phys. Rev. Lett.} \textbf{\bibinfo{volume}{90}},
  \bibinfo{pages}{095503} (\bibinfo{year}{2003}).

\end{thebibliography}

\end{document}